\documentclass[traditabstract]{aa}
\pdfoutput=1
\usepackage{graphicx}
\usepackage{txfonts}
\usepackage{layouts}
\usepackage{amssymb}
\usepackage{amsmath}

\usepackage{gensymb}
\usepackage[]{natbib}
\usepackage{natbib,twoopt}
\bibpunct{(}{)}{;}{a}{}{,} 
\makeatletter
\nonstopmode
\newcommandtwoopt{\citeads}[3][][]{\href{http://adsabs.harvard.edu/abs/#3}%
{\def\hyper@linkstart##1##2{}%
\let\hyper@linkend\@empty\citealp[#1][#2]{#3}}}
\newcommandtwoopt{\citepads}[3][][]{\href{http://adsabs.harvard.edu/abs/#3}%
{\def\hyper@linkstart##1##2{}%
\let\hyper@linkend\@empty\citep[#1][#2]{#3}}}
\newcommandtwoopt{\citetads}[3][][]{\href{http://adsabs.harvard.edu/abs/#3}%
{\def\hyper@linkstart##1##2{}%
\let\hyper@linkend\@empty\citet[#1][#2]{#3}}}
\newcommandtwoopt{\citeyearads}[3][][]%
{\href{http://adsabs.harvard.edu/abs/#3}
{\def\hyper@linkstart##1##2{}%
\let\hyper@linkend\@empty\citeyear[#1][#2]{#3}}}
\makeatother
\usepackage[colorlinks=true, linkcolor=blue, citecolor=blue,
urlcolor=blue]{hyperref}
\usepackage{caption}
\graphicspath{{Images/}}

\begin{document}

\title{Glimpses of stellar surfaces. I. Spot evolution and differential rotation of the planet host star Kepler-210}
\author{P. Ioannidis \& J.H.M.M. Schmitt}
\institute{Hamburger Sternwarte, Universit\"at Hamburg, Gojenbergsweg 112,
21029 Hamburg, Germany\\
\email{pioannidis@hs.uni-hamburg.de}}
\date{Received ...; accepted ...}
\abstract{We use high accuracy photometric data obtained 
with the {\it Kepler} satellite to monitor the
activity modulations of the Kepler-210 planet host star over {a time span of}
more than four years. Following the phenomenology of the star's light curve {in combination with} 
a five spot model, we identify six different so-called spot seasons. 
A characteristic, which is common in the majority of the seasons, is the persistent 
appearance of spots in a specific range of longitudes on the stellar surface.
The most prominent period of the observed activity modulations is different for each season 
and appears to evolve following a specific {pattern}, resembling the 
changes in the sunspot periods during the solar magnetic cycle. {Under the hypothesis that the
star exhibits solar-like differential rotation, we suggest} 
differential rotation values of Kepler-210 that are similar to or smaller than that of the 
Sun. 
Finally, we estimate spot life times between $\sim$60~days and $\sim$90~days, taking into 
consideration the evolution of the total covered stellar surface computed from our model.}

\titlerunning {Glimpses of stellar surfaces: Kepler-210}
\maketitle

\section{Introduction}

\begin{figure*}[t]
\includegraphics[width=\linewidth]{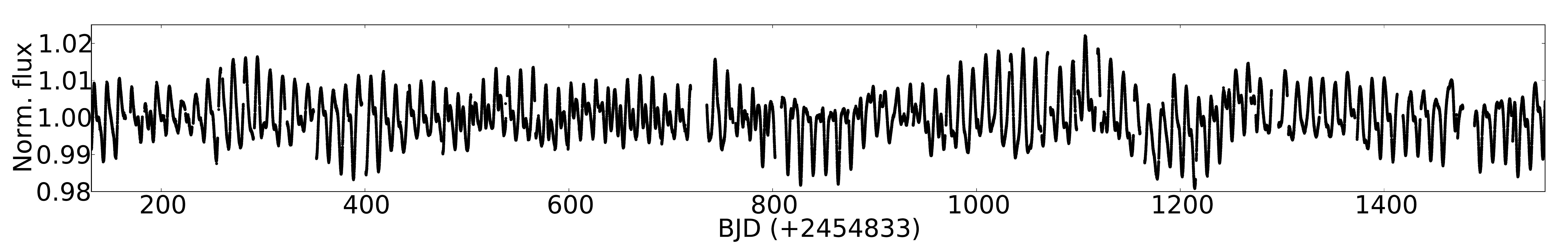}
\caption{The complete light curve of the Kepler-210 ($\sim$1$\,$400 days), normalized and
with the planetary transits excluded (see text for details).  }
\label{fig:lght}
\end{figure*}

The nature of the magnetic dynamo and the role of differential 
rotation in dwarf main-sequence stars is not yet fully understood. 
While these types of stars often exhibit substantial photometric variability, 
the amplitude of their surface differential rotation is expected to be small \citepads{2008JPhCS.118a2029K}. 
The inability of the small or absent differential rotation to organize the global magnetic field of those stars may result in the absence of activity
cycles as observed in the Sun \citepads{2006A&A...446.1027C}. 
As a result, the study of photospherically active stars and the measurement of their differential rotation rates is an
important piece of information for all dynamo theories. 

The rotational periods of stars can be measured with a variety of techniques 
including  monitoring of the {intensity variations of the cores of the
Ca~H+K lines}, 
spectral line broadening (for cases with known stellar radius and inclination), and the analysis of pseudo-periodic photometric modulations as a result of surface
inhomogeneities in the form of photospheric activity (spots). 
The successful operation of the space missions {\it CoRoT} \citepads{2006ESASP1306...33B} and {\it Kepler} \citepads{2010Sci...327..977B}, 
in combination with their {high photometric} accuracy and long, 
non-interrupted observations has revolutionized the studies of stellar rotation 
with period measurements being available for a very large number of field stars 
(\citeads{2013A&A...560A...4R}, \citeads{2014ApJS..211...24M}). Furthermore, 
the analysis of the photometric light curves of {\it CoRoT} and {\it Kepler,} using a variety of 
techniques, including power spectrum analysis \citepads{2013A&A...557A..11R}
and spot modeling (\citeads{2009A&A...506..263F}, \citeads{2009A&A...508..901H}, \citeads{2011A&A...532A..81F}, \citeads{2012A&A...547A..37B}),
makes it possible to measure the stellar differential rotation and other physical characteristics of star spots.

In this paper we use the light curve phenomenology of Kepler-210 
(i.e., the variations of the light curve between the stellar rotations) in 
combination with spot modeling and power spectrum analysis
to study its photospheric activity. 
In the first part of Sect.~\ref{sect:data} we describe the data and the properties of Kepler-210. 
In the second part of this section we present the details 
of the spot model used in our analysis and 
explain our choices regarding the number of free parameters in our model. 
In Sect.~\ref{sect:sp_pos}, we show the results of our combined analysis 
and, in Sect.~\ref{sect:ph_int}, we attempt to {provide a physical interpretation of
our results}. 
Finally, we conclude with a summary in Sect.~\ref{sect:conc}.

\section{Data and Analysis}\label{sect:data}

\subsection{Light curve and periodogram}\label{P_data}

The Kepler-210 system consists of a K dwarf with at 
least two planets orbiting around it \citepads{2014A&A...564A..33I}.
The {\it Kepler} light curve of the system was obtained from the 
STDADS\footnote{\href{http://archive.stsci.edu/kepler/data_search/search.php?
action=Search&ktc_kepler_id=7447200}
{http://archive.stsci.edu/kepler/data\_search/search.php?action=Search}
\href{http://archive.stsci.edu/kepler/data_search/search.php?
action=Search&ktc_kepler_id=7447200}{\&ktc\_kepler\_id=7447200}} 
archive and contains the long cadence data from quarters 
Q1 to Q17.  The removal of instrumental systematics from the {\it Kepler} light curves 
is quite cumbersome (\citeads{2012PASP..124.1073P}, \citeads{2012A&A...539A.137M}, 
\citeads{2012PASP..124..963K}), thus we {decided} to use the so-called 
corrected PDC-MAP data for our analysis. {\citetads{2012PASP..124..985S} provide a detailed description
of the philosophy behind the PDC-MAP approach and,  in our study
we rely on the ability
of this algorithm to remove instrumental effects, but to retain astrophysical 
effects.} 
In Figure~\ref{fig:lght}, we show the approximately 1$\,$400 days long light curve of Kepler-210 with 
each quarter normalized by its mean 
{\bf and} the planetary transits 
removed (using the parameters calculated by \citetads{2014A&A...564A..33I}). 
{The light curve shows} clear modulations with an amplitude of $\sim2\%$, similar to 
photospherically active stars, e.g., CoRoT-2 (cf., \citeads{2008A&A...482L..21A} and \citeads{2009A&A...508..901H}). 
To analyze those modulations we use the
generalized Lomb-Scargle (L-S) periodogram \citepads{2009A&A...496..577Z}
of the total light curve of Kepler-210. The peak with the highest power corresponds 
to a period of $P_{\star}=12.28$~days (see, Sect.~\ref{par:seasls}). However, we find 
evidence for additional power on various timescales which we would like to explore in
the following.

\subsection{Light curve modeling}

\subsubsection{Basics}
\label{par:meth}

To describe the relative stellar flux reduction $F_\mathrm{sp}/F_0$, which is due to the 
presence of an active region on the stellar surface,
we adopt a simple spot model, given by the expression
\begin{eqnarray}
\nonumber
\label{eq:spotflux}
\frac{F_\mathrm{sp}}{F_0} \, = \, 1 \, &-& \, \frac{S_\mathrm{sp} \cdot \cos[\,  \theta(t) 
]}{\pi R_{\star}^{2}} \,\\
&\times& \, ( 1 - t_\mathrm{sp}^4 )\cdot I[\, \theta(t)\, ,c_1,c_2] \ .
\end{eqnarray}
The terms $S_\mathrm{sp}$, $R_{\star}$, and $t_\mathrm{sp}$ 
denote the spotted surface, the stellar radius, and the relative spot
temperature (i.e., the ratio between spot temperature and photospheric temperature) respectively. The term $\theta(t)$ accounts for the angle between the line of sight towards 
the observer and the normal to the {spotted} surface at the time $t$ of the observation. 
The factor $I[\, \theta(t),c_1,c_2\, ]$ denotes the limb darkening (LD)
of the stellar disk, which depends on the angle $\theta(t)$. We 
use a quadratic LD parameterization, with the parameters $c_1$ and $c_2$ as limb-darkening coefficients (LDC).

\begin{figure}[t]
\includegraphics[width=\linewidth]{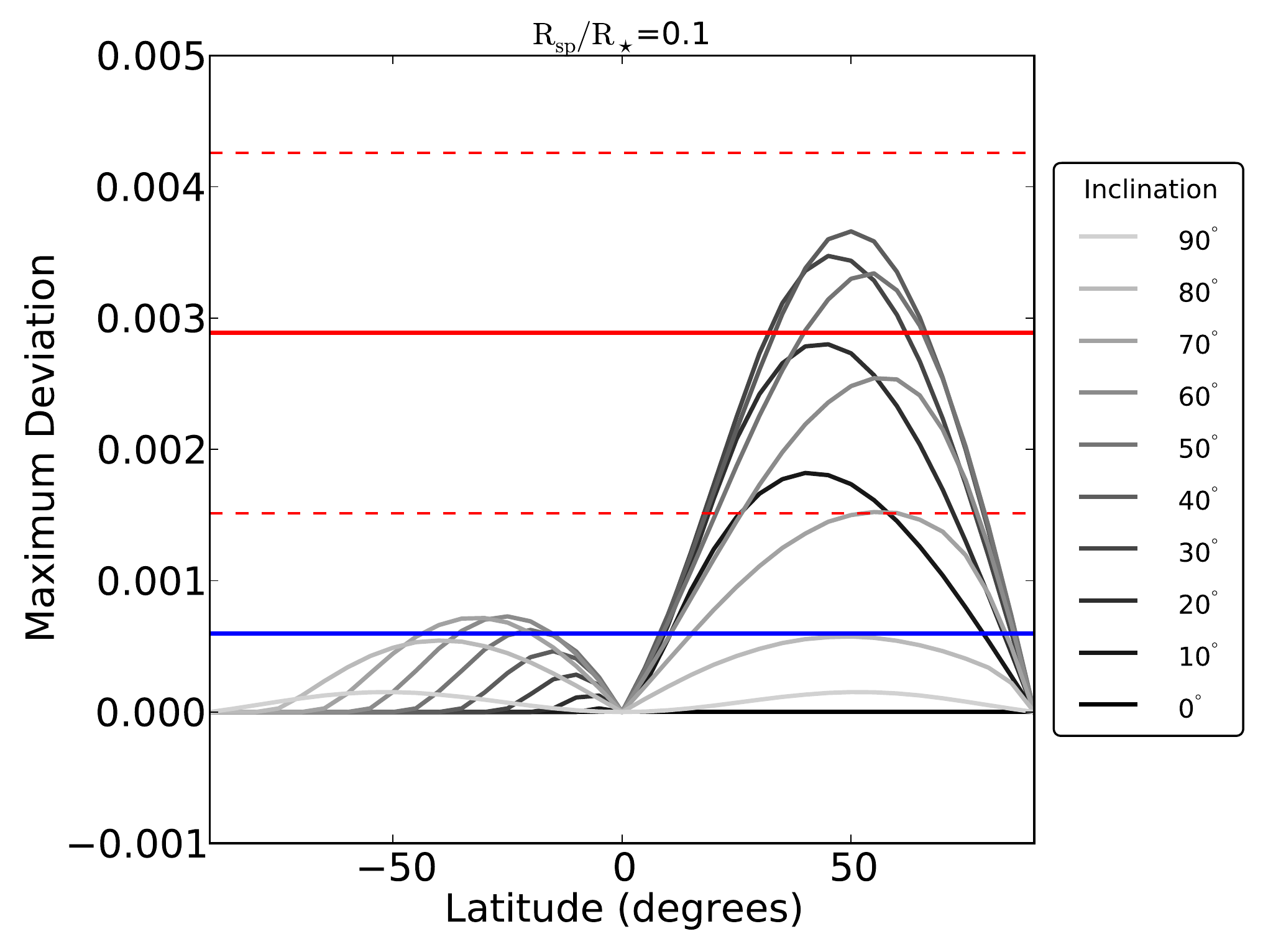}
\caption{Amplitude of the simulated light curves residuals after the  
subtraction of the fitted equatorial spot model. The red lines show the mean variation 
of the modulations from one rotation to the next (solid red line) and 
its uncertainty (dashed red lines). 
The blue solid line represents the level of the Gaussian noise of the Kepler-210 
light curve. }
\label{fig:lat_test}
\end{figure}

\noindent
For simplicity, we assume dark circular spots ($t_\mathrm{sp}$~=~0) and note that the
spot temperature can be calibrated later by increasing the values of the 
spot radius {(see discussion in Sect.~\ref{par:spot})}.
Owing to the fact that
Kepler-210 hosts a planetary system, we feel secure in assuming that the stellar
inclination is $\simeq$90${\degree}$ (\citeads{2012AJ....143...94T}, \citeads{2012A&A...541A.139F}, \citeads{2012ApJ...758...39J}, \citeads{2012ApJ...761...92F}, 
\citeads{2014ApJ...790..146F}, \citeads{2014ApJ...796...47M}), which we adopt in the following.

\subsubsection{Spot modeling}
\label{par:spot}
In our modeling, we assume {purely equatorial spots since} 
the latitude of an active region on the star can be regulated by its size $S_\mathrm{sp}$,
while the term $\theta(t)$ can be used to express the longitude of the region during the 
observation; $\lambda_\mathrm{o}=0{\degree}$ stands for the 
sub-observer point at the center of disk, with $\lambda_\mathrm{o}=-90{\degree}$ being 
the leading edge, and $\lambda_\mathrm{o}=90{\degree}$ the trailing edge of the disk.

To study the validity of this adopted equatorial spot model, we consider
noise-free light curves of fiducial spotted stars with various inclinations, where 
spots of some given size (we chose $R_\mathrm{sp}/R_\star$=0.1) 
are placed at various stellar latitudes.
The calculated light curves are fitted with an equatorial spot model, i.e., the spot latitude is set to zero, but the size of the spot is allowed to vary. 
We then compute the maximal (absolute) deviation (in amplitude) 
between the true light curve and the light curve modeled with equatorial 
spots and plot, in Fig.~\ref{fig:lat_test}, this maximal deviation as a function of spot latitude for various inclinations. 
As can be seen from Fig.~\ref{fig:lat_test}, these deviations 
disappear for latitudes of zero (where the spots are already on
the equator) and for polar spots, which produce essentially no rotational modulations.
The largest deviations occur for spots at latitudes between 30$\degree$-~60$\degree$ placed on stars seen with inclinations between 10$\degree$-~70$\degree$. In those cases 
the visibility duration of a spot does depend on latitude, while for inclinations near 90$\degree$ all spots have essentially the same visibility duration and spots at different latitudes differ only through the effects of differential limb darkening.

In Fig.~\ref{fig:lat_test} we also indicate the typical noise in the Kepler data of Kepler-210 (solid blue line)
as well as the observed mean peak to peak variation (solid red line) and its dispersion (dashed red lines).
Fig.~\ref{fig:lat_test} then indicates that, for inclinations in excess of about 70$\degree$, the 
noise in the data exceeds the maximally possible deviations introduced by the 
equatorial spot model. Since we assume an inclination close to 90$\degree$ for 
Kepler-210, we argue that our modeling approach with only equatorial spots  is in 
order and that reliable latitude information cannot be retrieved from the light curve.

\begin{figure}[t]
\includegraphics[width=\linewidth]{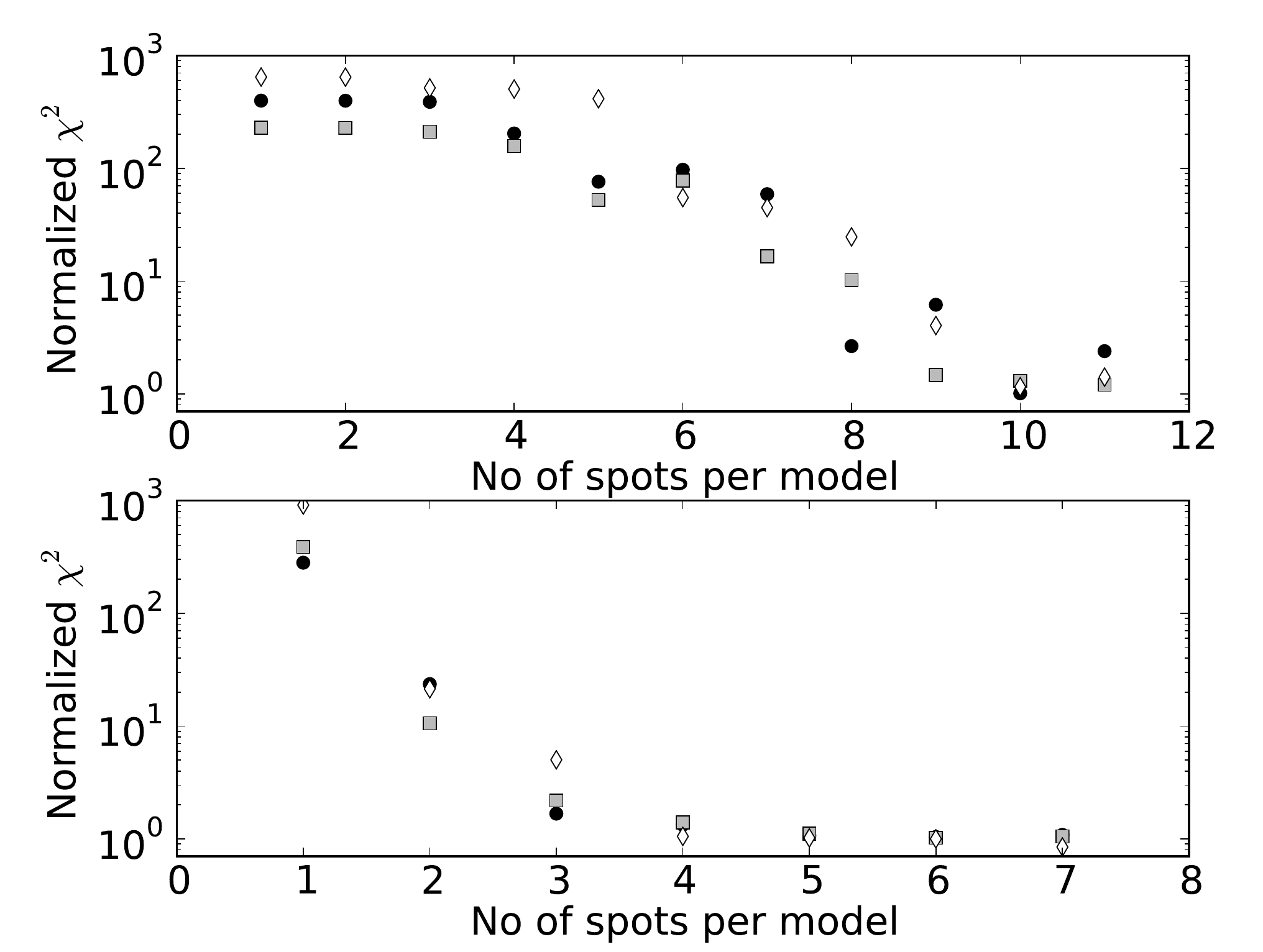}
\caption{$\chi^2$ goodness of the fit for models with different number of equally 
separated spots (upper panel) 
and free spots (lower panel). 
It is clear that the normalized $\chi^2$ is
approaching close to unity for models with at least
10 free parameters (see text for details).}
\label{fig:mod_comp_xi}
\end{figure}

\subsubsection{Light curve modeling}

\label{par:light}

To describe the observed light curve changes, we consider 
chunks of the overall light curve (see Fig.~\ref{fig:lght}),
equal to the leading period of the L-S periodogram, 
separated by a quarter of that period, i.e.,
the individual light curves are not independent. 
The light curve modulations are not sufficiently stable from one rotation to
the next, so none of the light curve chunks are
identical. To cope with this problem we use {equatorial spot models to describe the observed modulations}. Each spot model then
represents a unique passage of this region over the visible 
hemisphere of the star.

{It is well known that the
problem of modeling the spot distribution on a two-dimensional surface into
a one-dimensional light curve is ill-posed, i.e., in general there is more than 
one model to appropriately describe the observed light curve modulations.
For our light curve modeling, we consider two different approaches:  
in the first approach, we consider $N$
circular dark spots, where the $i^{th}$ spot has a radius $R_\mathrm{i}$ and is located 
on the equator at some longitude $\lambda_\mathrm{i}$; as a second approach, we again consider
 $N$ circular dark spots, this time however, at fixed, equidistant locations
around the stellar equator.  Since the star 
appears to  always be covered by  spots and since stitching of the 
individual quarters may not be fully correct,
it is impossible to know the correct normalization of the light curve. As a result, we include an estimate of the level 
of the unspotted flux in our model.  To assess the goodness of fit, 
we compute the $\chi^2$-values of our fit for all models.
Although the quality of the model fits increases with the number of spots, i.e., 
the number of the available free parameters, this number should be as low as 
necessary.  Therefore we first address the question of the effective number of
free parameters in our problem.}

In the upper panel of Fig.~\ref{fig:mod_comp_xi}, we first consider the $\chi^2$-values for models
with a given number of fixed spots at equidistant locations around the equator for three 
randomly chosen light curve chunks (described by different symbols).  
As the spots {are described by} only one free parameter (their respective radius), the number of spots is equal to the 
number of free parameters of the model plus one (for the normalization factor).  
It is clear that around ten such spots are required to obtain acceptable fits.

\begin{figure}[t]
\includegraphics[width=\linewidth]{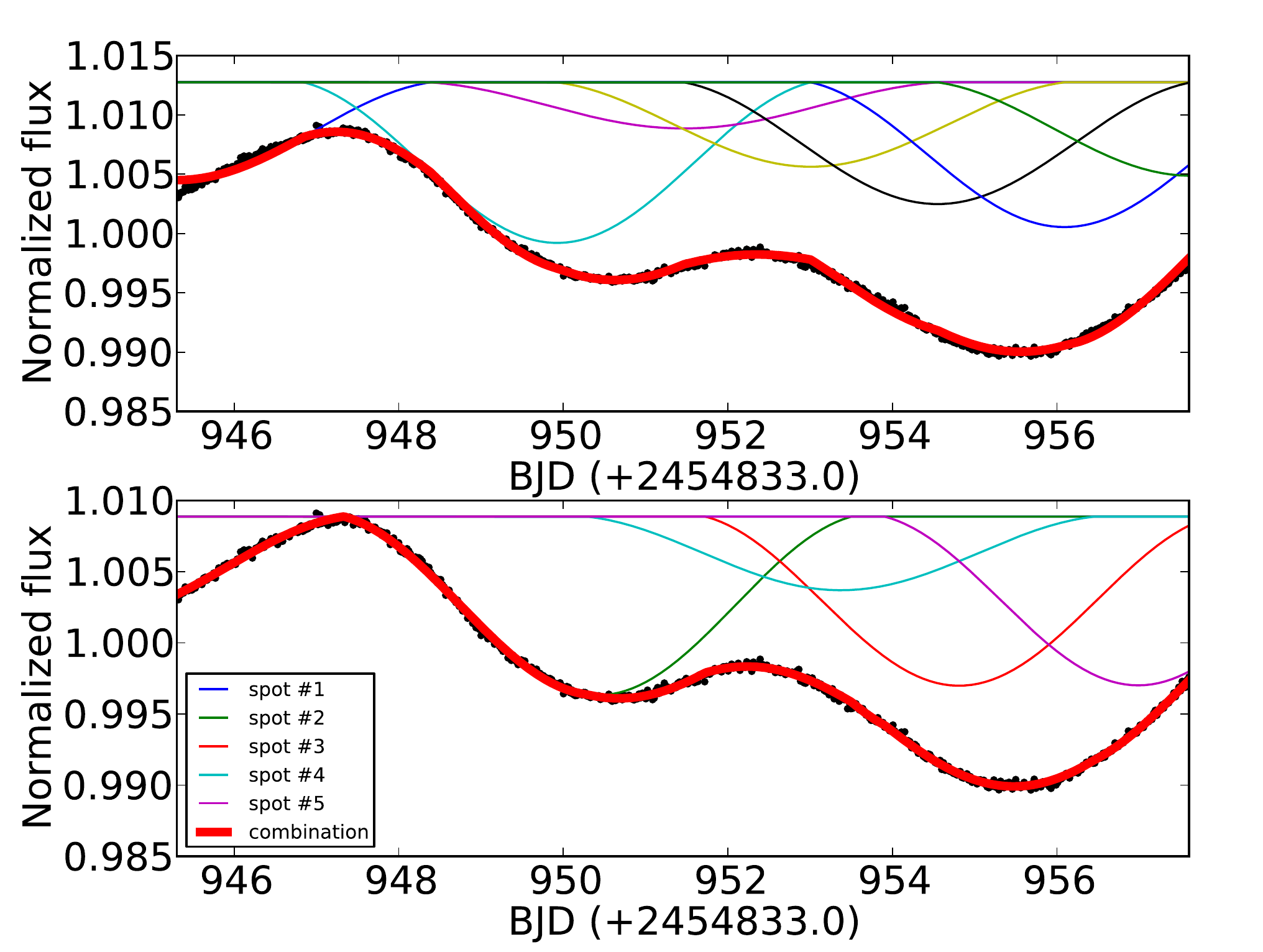}
\caption{Top: Visual representation of the 10 fixed spot model. 
Each of the thin colored lines represents the flux reduction caused from 
each of the spots during a full rotation. 
The combined flux reduction is given from the aggregation of the
five spots models (thick red line). Bottom: As for top panel, but for the five-spot model.}
\label{fig:model}
\end{figure}

In the lower panel of Fig.~\ref{fig:mod_comp_xi} we consider models with a variable number of spots at variable positions along the equator and plot again 
the $\chi^2$-values as a function of
spot number.  Clearly, Fig.~\ref{fig:mod_comp_xi} suggests that one needs about 
five free spots to describe the observed modulations appropriately.
A typical fit to one of our light curves using 10 equidistant spots is shown in the top
panel of Fig.~\ref{fig:model}. We note that each colored curve represent the 
passing of one spot from the visible hemisphere of the star while the thick red line
shows their sum, i.e., the actual model fit of the light curve.  The same
light curve fitted with a five spot model with variable positions is shown in the
lower panel of Fig.~\ref{fig:model}. We can immediately identify the models for 
spots \#3, \#4, and \#5. Spot \#1 cannot be
 seen since it is covered under the combined model (thick red line). It is
obvious that without spot \#1, which shares the same characteristics 
as spot \#4, the section of the light curve with $<$947.5 BJD (+2454833.0)
would have been impossible to model.  We therefore conclude that one needs about
ten free parameters to arrive at an adequate fit to the observed light curves. 

The total light curve is affected by data drop-outs in various parts. As a result the
number of data available, points fluctuate from one light curve 
chunk to the next. To avoid errors in our calculations, we model only those light curve chunks with a phase coverage larger than 70\%. 
The final number of analyzed chunks is thus 379 out of 462. Two adjacent light curve chunks
are obviously correlated since they share 75\% of their data points, hence the
estimate of the size and longitude for each spot is done more than once.
We finally use a MCMC (Markov-Chain Monte-Carlo) approach 
to estimate the size and the longitude of each spot, as well as the errors thereof.

\begin{figure*}[tp]
\centering
\includegraphics[width=17cm]{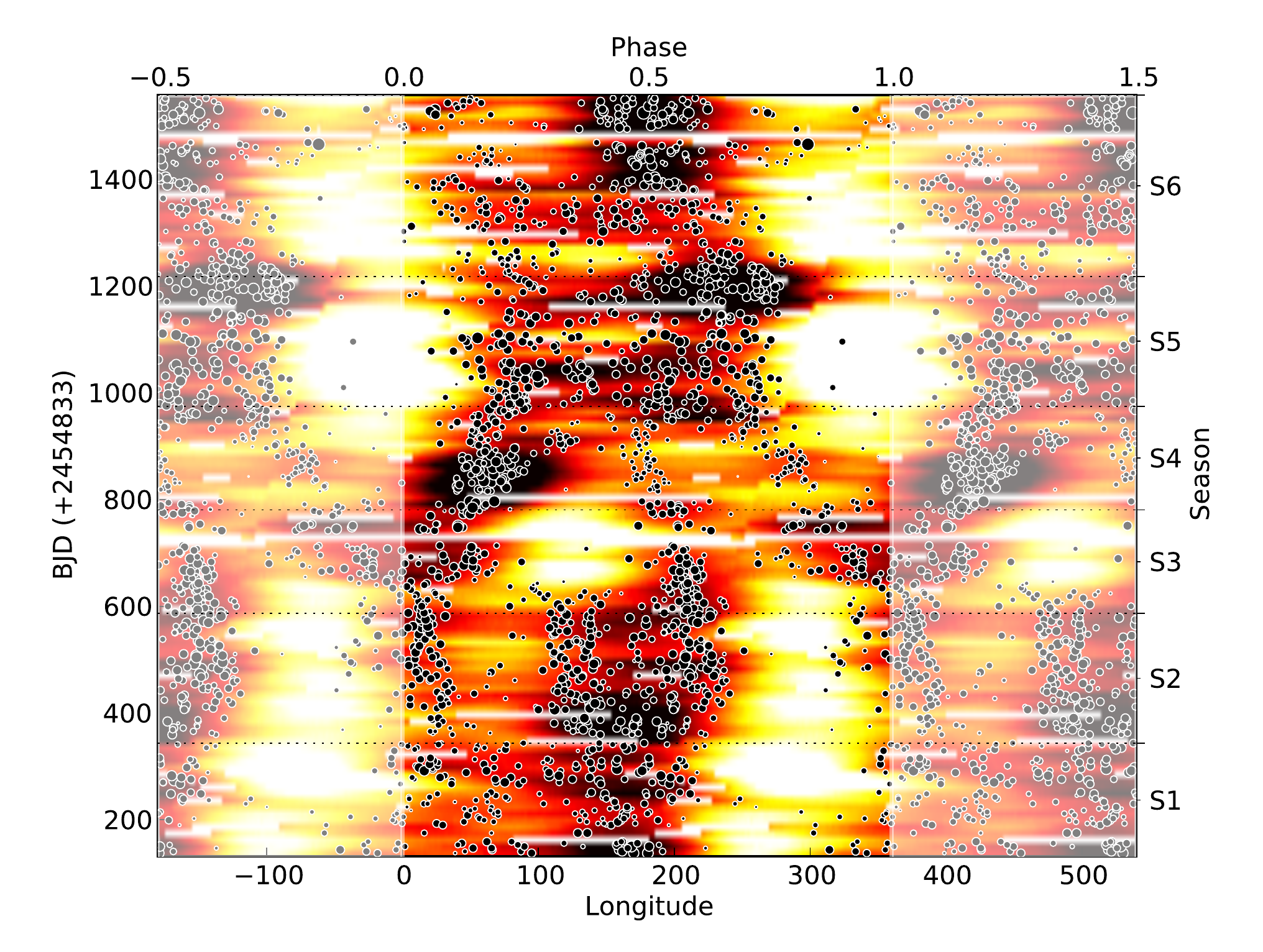}
\caption{The phase folded data, with  the calculated spot
longitudes over-plotted. The plot is extended a half phase in each direction to become
more readable (shaded areas).}
\label{fig:res}
\end{figure*}

\section{Results}\label{sect:sp_pos}
\subsection{Spot configurations and stellar spot seasons}\label{sect:phenom}

Since we are interested in localizing the spots on the stellar surface, we choose 
a five-spot model with variable positions and radii in the following.
In Fig.~\ref{fig:res} we show the results of our spot analysis for Kepler-210 by
plotting the calculated {- from the model -} positions of the modeled spots 
vs. time (on the y-axis).
The color-coded plot represents the phase-folded light curve,
and the calculated spot longitudes are over-plotted at the appropriate spot
longitudes.
The size of each dot 
corresponds to the calculated size of the spot.
The star spot
sizes and longitudes show some interesting features in the behavior of the
spots on Kepler-210. In the following we refer to time as the number of days
elapsed since the date BJD~=~2454833.0.

At the beginning of the Kepler observations the activity on Kepler-210 is concentrated into two regions {centered} between 0${\degree}$-20${\degree}$ longitude and 180${\degree}$-200${\degree}$ longitude. As a result one hemisphere is very active, i.e., the one between 0${\degree}$-180${\degree}$, while the other half of the star remains almost spotless.  

By day $\sim$ 350, the region at 180${\degree}$ appears to split up in two parts, one apparently moving towards smaller longitudes and the other towards larger longitudes. By day $\sim$630, the region moving towards smaller longitudes disappears and the region remains spotless for about 200 days. At the same time activity appears at longitude 200${\degree}$-300${\degree}$, i.e., a region that was found inactive before. Also, the activity region at 0${\degree}$-20${\degree}$ starts drifting towards larger longitudes. By day $\sim$800, the activity in the longitude range 200${\degree}$-300${\degree}$ disappears again and that range 
remains more or less spotless for the rest of the Kepler-210 observations. Between 800~days and 1 000~days, three activity concentrations are visible, two at about 200${\degree}$ and 300${\degree}$ longitude moving towards smaller longitudes and the already mentioned region at 0${\degree}$-20${\degree}$, which is now moving towards larger longitudes. Between days 1 000 and 1 250, the activity appears 
to be concentrated more and less homogeneously between longitudes 50${\degree}$-250${\degree}$. By around day 1$\,$200, the activity near longitude 120${\degree}$ starts thinning out and the recognized two activity complexes are drifting towards smaller longitudes. Based on this phenomenology, we divide the data of Kepler-210 into six seasons S1 .... S6: S1:~130~days~-~350~days, S2:~350~days~-~600~days, S3:~600~days~-~800~days, S4:~800~days~-~1000~days, S5:~1000~days~-~1250~days and  S6:~1250~days~-~1600~days.

\begin{figure}[t]
\includegraphics[width=\linewidth]{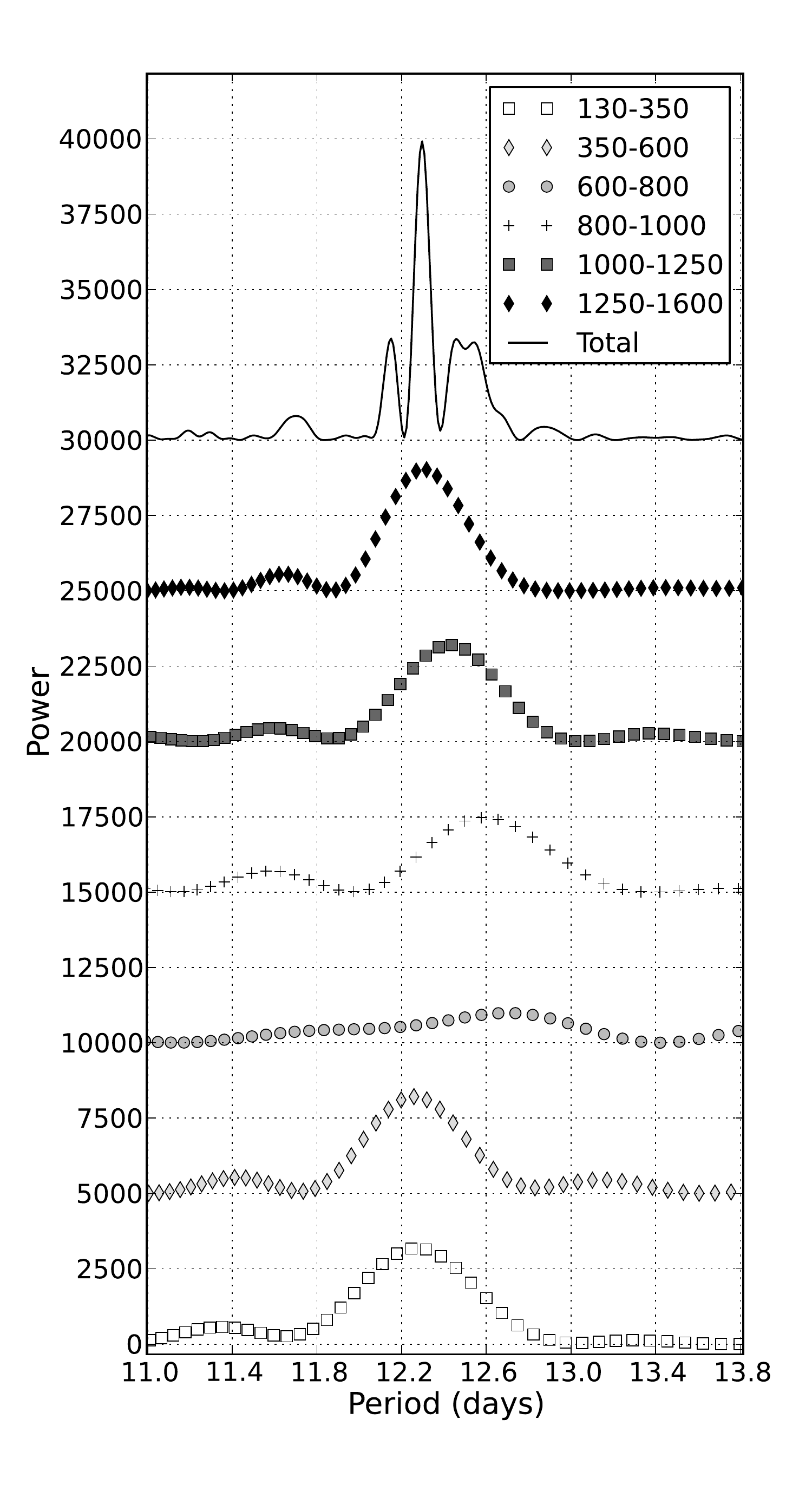}
\caption{L-S periodograms of each so-called stellar season. The solid line represents the
total L-S periodogram of the light curve. The transition from small periods to longer and
then gradually to small again resembles the butterfly diagram of the Sun (see Sects~\ref{par:seasls}~\&~\ref{par:diffrotkep210}).}
\label{fig:ls-seasons}
\end{figure}
\begin{figure}[t]
\includegraphics[width=\linewidth]{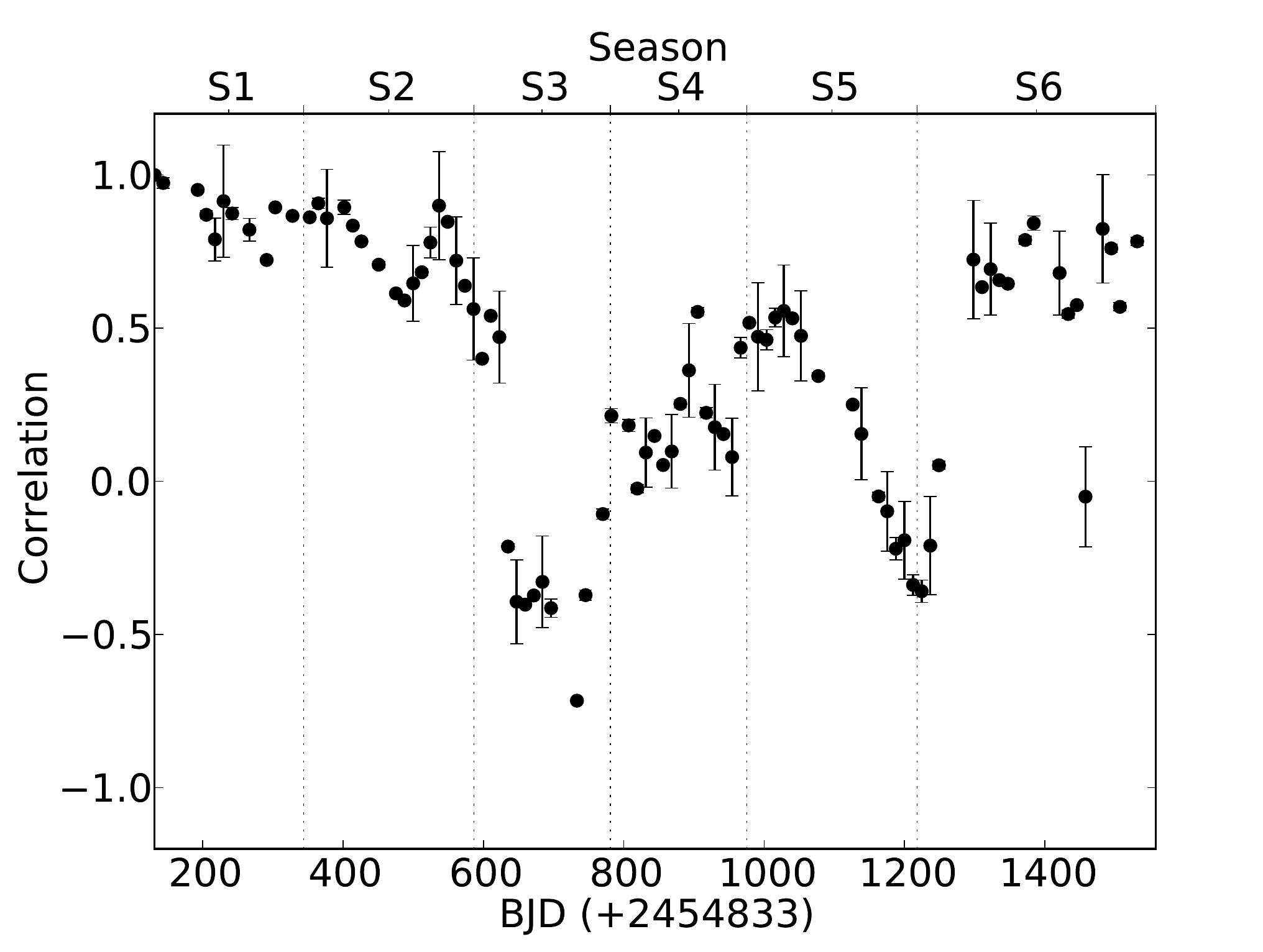}
\caption{The cross-correlation between the different phases with period $P$ = 12.28 
days. The outlying points in times $\sim$750 days and $\sim$1450 days originate
to anomalies in the light curve due to the transition from one {\it Kepler} observation quarter to the next.  }
\label{fig:croscor}
\end{figure}

\subsection{Power spectrum analysis}
\label{par:seasls}

The introduction of the so-called seasons in Sect.~\ref{sect:phenom} was performed
purely phenomenologically. 
We now calculate the L-S periodogram for each season (shown in Fig.~\ref{fig:ls-seasons}) and note that the value of the maximum power periods 
varies from season to season, with exception of seasons S1 and S2. The L-S of season S3 has the smallest power, owing to the existence of multiple periodicities and the maximum power of this season appears to be at periods slightly larger than 12.6 days. Starting from season S3 and up to S6, there is an obvious diminution of the prominent period value from $\sim$12.6 days to $\sim$12.3 days. In general, the 
period range which we calculate for the different seasons explains the structure of the L-S diagram of the total light curve (continuous line), with exception of the peak at $\sim$12.1 days, the origins of which we are not able to clarify.

\subsection{Cross-correlation between the different light curve portions}

Another method adequate to quantify any differences between the observed seasons is to compare the evolution of the light curve modulations in time. To that end, we separate the light curve in parts (phases) equal to the leading period of the L-S periodogram (i.e., 12.28~days) following the expression

\begin{equation}
\Phi_\mathrm{i}\,=\,[t_{0}+P \cdot \mathrm{i}\,,\,t_{0}+P\cdot (\mathrm{i}+1)]\quad \mathrm{with}\quad\mathrm{i}=0,1,2...N,
\end{equation}
where $P$, $T_{0}$, and $N$ denote the period, the time of the first observation, and the total number of phases respectively. 
In Fig.~\ref{fig:croscor}, we show the cross-correlation between the different phases $\Phi_\mathrm{i}$ of the light curve and the  phase $\Phi_\mathrm{0}$ (reference phase). The error bars are inversely proportional to the common number of points per phase pair, i.e, zero error denotes equal number of points between the reference and the examined phase. In the same fashion as with the L-S periodogram, the seasons are also easily distinguishable in Fig.~\ref{fig:croscor}, marked by rapid changes in the cross-correlation between the phases.

\section{Physical interpretation}\label{sect:ph_int}

So far we have shown that the rotational periods of the active regions of Kepler-210 appear 
to vary in time. But what is the physical interpretation of the results of Sect.~\ref{sect:sp_pos}?

\subsection{Active longitudes}

There is a clear preference {for some} longitudes where spots {prefer} to appear,
while other surface areas remain more or less spotless.
Using our algorithm, we find that there is an area 
between 150$\degree$ and 200$\degree$ in longitude, which is covered
by spots in every observed rotation of the star. On the 
other hand, about 100 degrees in longitude are covered by spots 
 during ~only 20\% of the total observation
time.

\subsection{Differential rotation}
\label{par:diffrotkep210}
As discussed in Sect.~\ref{sect:sp_pos}, the leading period of the activity modulations
of each season is different (see Fig.~\ref{fig:ls-seasons}). Thus, in agreement 
with the claims of \citetads{2013A&A...560A...4R}, 
we interpret those changes as evidence for differential rotation,
{now assuming that the observed difference in period can be attributed to a difference in 
latitude. We specifically assume a solar-like differential rotation pattern with shorter rotation periods occurring near the equator. 
Furthermore, we can estimate the magnitude of the differential rotation as a function of the assumed spot latitude,
assuming the same analytical form of the 
differential rotation law as applicable for 
the Sun:
\begin{equation}\label{eq:difrotsun}
\Omega_{\mathrm{obs}}(\phi) = \Omega_\mathrm{eq} \cdot(1 - \alpha \times sin^2\phi)  
,\end{equation}
with $\phi$, $\Omega_{\mathrm{obs}}$, and $\Omega_\mathrm{eq}$ denoting the latitude and the rotation
rates at given latitudes and the equator, respectively.}
Our argument is 
also supported  by the visible drifts of active regions in Fig.~\ref{fig:res}; the spot groups with larger periods 
than the phase-folding period (P = 12.28, see Fig.~\ref{fig:ls-seasons}),  appear with a small delay from one 
phase to the next, while the spot groups with shorter periods appear earlier. 
A nice example of these motions can be found in season S4, between
800 and 1000 days. 

A characteristic of the Fig.~\ref{fig:ls-seasons} worth mentioning is the so-called jump of the spots from small latitudes (season S2) to larger latitudes (season S3), which is then followed by a smooth migration of the spots back to smaller latitudes. This behavior resembles the behavior of sun spots as they appear in the 
famous  butterfly diagram 
for the Sun.  During an 11-year magnetic cycle, the latitudes 
of the appearing sunspots first increase rapidly and then, gradually, move 
closer to the equator.

Given the fact that the peak to peak amplitude of the modulations is not 
reduced dramatically during season S3, we assume that the maximum
latitude $\phi_\mathrm{max}$ of the spots ought to {remain not too far away
from the stellar equator},  so that the 
reduction of the spotted areas (due to their projection)  remains insignificant. 
To test {this hypothesis we calculate the strength of the differential 
rotation $\alpha$} as a function of the stellar latitude $\phi$
using the equation

\begin{figure}[t]
\includegraphics[width=\linewidth]{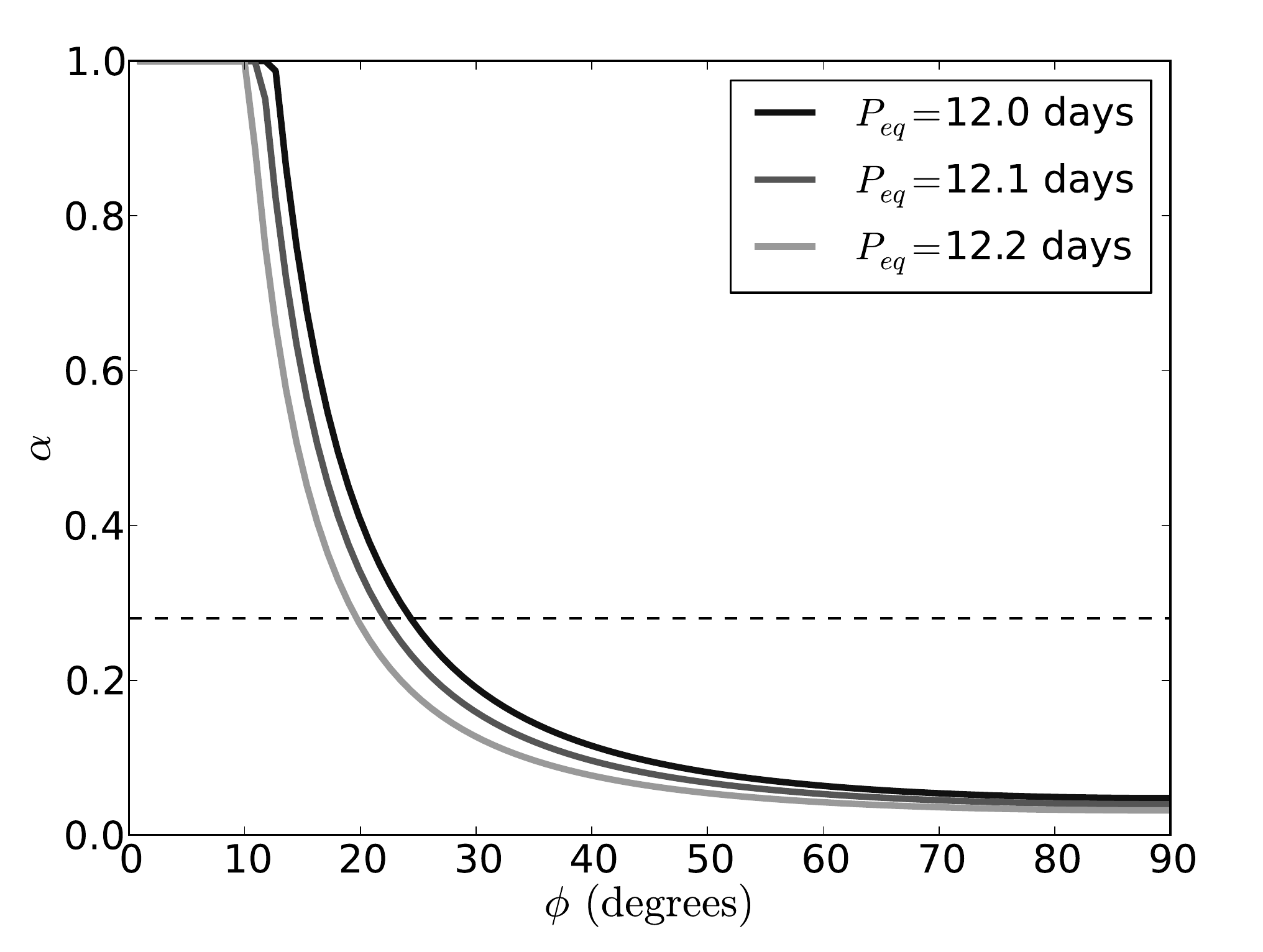}
\caption{Estimation of the differential rotation $\alpha$  of 
the star, assuming spots with rotational
period $P_\mathrm{sp}\simeq12.6\,\mathrm{days}$ for latitudes 
between $0\degree$ and $90\degree$. The dashed line indicates
the value of the differential rotation of the Sun.}
\label{fig:difrot}
\end{figure}
\noindent

\begin{equation}
\alpha = \left( 1 - {P_\mathrm{eq}\over P_\mathrm{obs}(\phi)}\right)\cdot{1\over {\sin^{2} \phi}},
\end{equation}
where $P_\mathrm{obs}(\phi)$ is the rotational period at a 
given stellar latitude $\phi$ and
$P_\mathrm{eq}$ denotes the equatorial rotational period. 
In Fig.~\ref{fig:difrot}, we plot 
our estimates for the differential rotation 
strength $\alpha,$ assuming different stellar latitudes $\phi$ for spots with rotational 
period 12.6~days (i.e., the spots which are assumed to be
responsible for the peak close to $\sim$12.6~days 
in the L-S of season S3), given three different values for the
equatorial rotation period of the star. The dashed line in the same
diagram indicates the value of the solar differential rotation, i.e., $\sim0.28$ (\citeads{1970SoPh...12...23H}, \citeads{1990ApJ...351..309S}). 

{At this stage we cannot provide a direct connection between the observed rotational period of the
spots and their latitudes.
However, we can use our estimate for the strength of the differential rotation in Fig.~\ref{fig:difrot} to put some constraints on the possible latitudes of the spots with 
rotational period $P_\mathrm{sp}\simeq12.6\,\mathrm{days}$.}

{In Fig.~\ref{fig:difrot}, we see that the strength of the required differential rotation 
reaches values higher than that of the Sun, assuming that the latitudes of the spots with 
rotational period $P_\mathrm{sp}\simeq12.6\,\mathrm{days}$ have 
values lower than $\sim$ 25$\degree$. As a result, we do not favor this assumption
since it might not be easily explained by theory \citepads[e.g.,][]{2008JPhCS.118a2029K}}

{Furthermore, based on the assumption upon which we observe the star equator, the
hypothesis that the spots of season S3 occupy latitudes larger than $\sim$40$\degree$
would require dramatic variations in the average spot sizes from one season to the next to maintain the relatively stable amplitude of $\sim$2\%, which we observe in the light
curve of Kepler-210. We note that while the transits of Kepler-210 b are affected by 
spot-crossing events, the estimated size of those spots is not sufficient to produce the 
observed modulations.}

{Consequently, we suggest that the spots of season S3
with rotational period $P_\mathrm{sp}\simeq12.6$~days ought to have latitudes in the 
range of between $25\degree  \lesssim\phi_\mathrm{max}\lesssim40\degree$. Incidentally, this 
is also the latitude range where we observe the sun spots with the highest latitudes, i.e., during the solar maximum. 
Following the hypothesis that the latitudes of the spots with rotational period $P_\mathrm{sp}\simeq12.6\,\mathrm{days}$ 
lay in the range $25\degree  \lesssim\phi_\mathrm{max}\lesssim40\degree$, 
the differential rotation of the star is similar and slightly smaller in 
comparison to the differential rotation of the Sun}.

During seasons S3 and S4 another abnormal event takes place: It is the only time during the 
Kepler-210 observations when the area between $\sim$200${\degree}$ and 
$\sim$360${\degree}$ is covered by spots. At the same time, the rest of the usually 
spotted stellar surface becomes free of spots. Although
the connection of this event to a physical process might not be trivial, it is an evidence 
that a dramatic change happened to the star during that season (see, Fig.~\ref{fig:res}).

\subsection{Spot formation and spot life times}
The observed flux variations of Kepler-210 are similar from one rotation to the
next (see Figs~\ref{fig:croscor}). The latter fact suggests that
the life time $\tau_\mathrm{sp}$ of the active regions
is much longer than the stellar rotation period $P_{\star}$.

In Fig.~\ref{fig:lght},we observe parts of the
light curve where the peak-to-peak amplitude of the modulations is larger. 
Owing to the lack of a plateaued part in the light curve, we conclude that 
there is no time during the {\it Kepler} observations of Kepler-210 with a
complete absence of spots on the visible hemisphere of the star. As a result, we assume that 
the short-term variations of the total spot coverage are the result of 
simultaneous occupations of stellar longitudes closer than the half visible hemisphere
from different spot groups.

\begin{figure}[t]
\includegraphics[width=\linewidth]{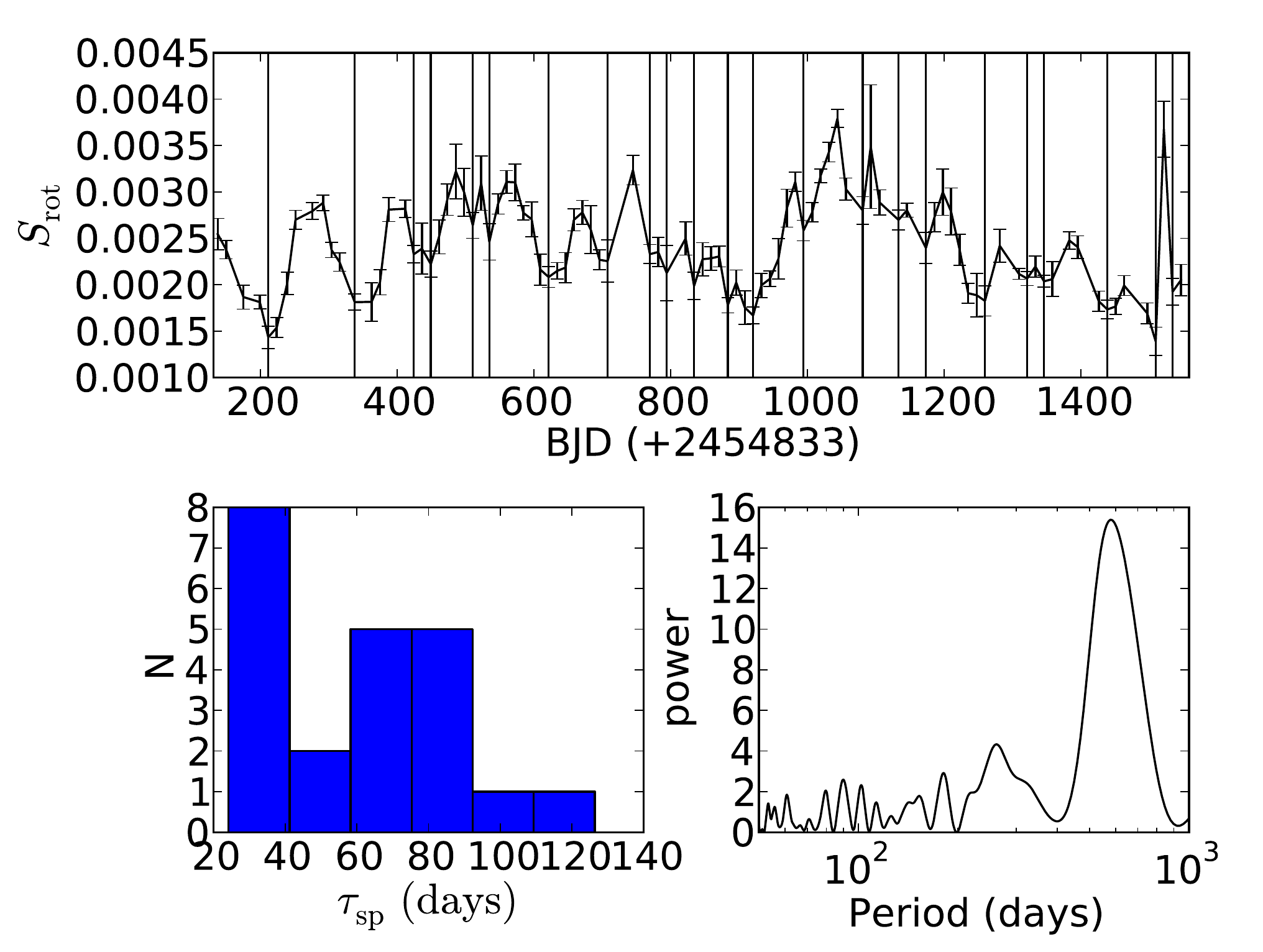}
\caption{Top: Evolution of the size of the total spotted area, calculated as the
summed area of the spots, which define the model for each light curve chunk. 
The vertical lines indicate the local minima of the curve.
Bottom-left: The distribution of the spot life times (see text for details).  
Bottom-right: The L-S periodogram for the evolution of the spotted area size.}
\label{fig:lifet}
\end{figure}
\noindent
There are two possibilities for such an association to occur; either a spot appears
in an area with a neighboring longitude to a preexisting spot, or two spots
with different stellar latitudes ``meet'' due to their differential rotation. From a 
visual examination of Fig.~\ref{fig:res}, we observe that both effects are 
taking place, with the former effect being more dominant. Therefore we suggest 
that it is possible to estimate the life time of the spots by measuring
the duration of the simultaneous appearance of spot groups in the 
same stellar hemisphere.

Using the estimated spot radii we can calculate
the total covered area ($S_\mathrm{rot}$) for each of the 
light curve chunks (see, Sect.~\ref{par:meth} for details).
The top panel of Fig.~\ref{fig:lifet} shows 
the evolution of the total covered area
as a function of time. As the duration of each fluctuation, we consider
the range between two local minima of the covered 
area curve, which are marked with vertical lines in Fig.~\ref{fig:lifet}.

In the lower left panel of Fig.~\ref{fig:lifet}, we show the 
suggested spot lifetime distribution, which ranges between
$\sim$25~days and $\sim$130~days. There are two
maxima, one close to $\sim$25~days and an other
around $\sim$90~days. After visual examination 
of Fig.~\ref{fig:res}, we can
confirm that the duration of 
the dark features (spotted areas) varies 
between $\sim$60~days and $\sim$90~days.
The peak close to $\sim$25~days  (see, Fig.~\ref{fig:lifet}) is
probably a bias caused by the detection of consecutive local
extrema in the curve of the total covered area.

The L-S periodogram of the spotted area, displayed
in the lower right panel of Fig.~\ref{fig:lifet}, shows several
low significance peaks for periods in the range between
$\sim$60~days and $\sim$90~days. The most significant
peak is found for a period around $\sim$600 days. The
long term variation responsible for this peak is
visible in the upper panel of Fig.~\ref{fig:lifet}, where 
the total spotted area appears to become larger close to 
the end of season S2, then drops for seasons S3 and S4,
and rises again in season S5.

\section{Conclusions}\label{sect:conc}

Using the phenomenology of the Kepler-210 light curve in combination
with the results of a five-spot model, we 
study the behavior of the spotted areas on the star
(i.e., their relative periods and the longitudes at which they appear)
and their changes in time (see Fig.~\ref{fig:res}). Based
on the spot phenomenology we identify six different ``spot seasons'' and
demonstrate that there are differences in the dominant periods of the 
L-S periodograms corresponding to each season (see Fig.~\ref{fig:ls-seasons}). Additionally we show that
the seasons also manifest themselves  as differences in the correlation between
the corresponding parts of the light curve (see Fig.~\ref{fig:croscor}).

According to Fig.~\ref{fig:ls-seasons}, the relative period of spots in
the subsequent seasons appears to change in the same fashion
as the relative rotational period of sunspots during the solar cycle, i.e., 
the relative starspot period appears to change from lower to higher values between seasons S2 and S3,
while it diminishes gradually from season S4 until the end of the {\it Kepler} observations. 
A common characteristic between all seasons, with the exception of seasons S3 and S4, is the  persistent appearance of the spots in a specific longitude range of the star.

Assuming solar-like differential rotation we show that the
value of the strength of the differential rotation $\alpha$ {ought to be similar or lower 
to that of the Sun under the hypothesis that the spots with the higher periods 
have latitudes in the range of between $25\degree \lesssim\phi_\mathrm{max}\lesssim40\degree$
(see Fig.~\ref{fig:difrot}).}
Furthermore, we estimate the spot life times
using the radii of the spots that were computed with our model fit.
As a result, we conclude that the spot life times of Kepler-210 vary 
between $\sim$60~days and $\sim$90~days (see Figs~\ref{fig:res} \&~\ref{fig:lifet}).

The behavior of the active regions on the photosphere of Kepler-210 (i.e., the shift from small asterographic latitudes to higher and vice versa) is comparable to the migration of sunspots during an 11-year solar magnetic cycle (see Fig.~\ref{fig:ls-seasons}). We estimate that the duration of this phenomenon on Kepler-210 is similar, or somewhat longer, than the total {\it Kepler} observation time (i.e., $\sim$4 years), however, additional long-term observations
are clearly needed to check whether this behavior of Kepler-210
as observed by {\it Kepler} is actually periodic and indeed the
result of a magnetic cycle. 

\section{Acknowledgments}

PI acknowledges funding through the DFG grant RTG
1351/2 Extrasolar planets and their host 
stars.

\bibliographystyle{aa}
\bibliography{references}

\end{document}